# Ensuring Responsible Outcomes from Technology

Aaditeshwar Seth
IIT Delhi and Gram Vaani
India
aseth@cse.iitd.ac.in, aseth@gramvaani.org

*Abstract*—We attempt to make two arguments in this essay. First, through a case study of a mobile phone based voice-media service we have been running in rural central India for more than six years, we describe several implementation complexities we had to navigate towards realizing our intended vision of bringing social development through technology. Most of these complexities arose in the interface of our technology with society, and we argue that even other technology providers can create similar processes to manage this socio-technological interface and ensure intended outcomes from their technology use. We then build our second argument about how to ensure that the organizations behind both market driven technologies and those technologies that are adopted by the state, pay due attention towards responsibly managing the socio-technological interface of their innovations. We advocate for the technology engineers and researchers who work within these organizations, to take up the responsibility and ensure that their labour leads to making the world a better place especially for the poor and marginalized. We outline possible governance structures that can give more voice to the technology developers to push their organizations towards ensuring that responsible outcomes emerge from their technology. We note that the examples we use to build our arguments are limited to contemporary information and communication technology (ICT) platforms used directly by end-users to share content with one another, and hence our argument may not generalize to other ICTs in a straightforward manner.

*Keywords—ICTD, community media, social development, technology appropriation, inclusion, exclusion, inequality, political economy, codetermination*

## I. Introduction

There is growing concern that although rapid technology development has produced amazing outcomes, there have also been significant harmful effects for many reasons. Technology providers may be unable to control what the technology gets used for and by whom, they may not understand the limitations of their own technology, the regulatory response of the state might be too slow, and the technology may increase inequalities by servicing the rich and privileged much more than the poor and marginalized. Even technologies specifically designed to address certain social development objectives and to reduce inequalities, may fail to do so.

In this context, we try to address two goals. First, we list out several reasons because of which technologies may fail to produce the intended social development impact. We do this by using as a case-study our own experiences of having built and operated a voice-based community media service in rural India for several years. We understand this initiative most intimately as compared to any other systems we have studied in the research literature, and therefore we are able to explain in detail many aspects that we had to manage beyond just the technology to bring impact via this service. Second, we distil out specific processes we developed to manage the technology and discuss how they can be incorporated in other technological initiatives, problems that arise in these initiatives when the processes are missing, and how can the organizations which operate these initiatives be persuaded to build such processes that may require them to go beyond their core task of technology engineering alone.

We begin with giving a background of the community media service operated by us, then describe the challenges we faced and how we solved them, followed by a discussion of other market and state led technology initiatives about what is missing in their operations. We also raise an appeal for technology researchers and engineers to understand more about how the technologies produced by them are used in the world, ensure that their organizations do more to manage the use of technology, and work towards bringing social development through their technology.

## II. Background of Mobile Vaani

The concept of community media is centred in the idea that the needs of local or interest based communities is best understood by the community itself, and editorially driven mass media run by external institutions is not able to address these needs largely because of their lack of local context. The community itself should be able to define the agenda of a media meant for it, create its own content, and distribute the content to the community. Several such models for community embedded media initiatives exist in India itself. *Khabar Lahariya* runs local chapters of rural newspapers written and edited by local women, with a feminist perspective [1]. *Video Volunteers* trains community correspondents across the country to produce videos about human rights and governance issues in their communities [2]. Several community radio stations espouse the same perspective and are operated by people from local communities who create radio programmes and raise funds to sustain the station [3]. In this rich mix of initiatives, we started a mobile phone based community media service in rural central India in 2012, which works using Interactive Voice Response (IVR) systems to enable a voice driven communication platform for communities. Our choice to build a voice-based platform was driven by low literacy in the communities where we wanted to work, for whom voice would be more easily accessible as compared to text. Our choice to use IVR systems was driven by the rapid growth of mobile phone coverage, where even if people did not use the Internet due to challenges of affordability or digital literacy, they would still be able to use phones to make calls and press buttons to navigate content. Our conviction was also shaped with our prior experience of having built a community radio automation system with an inbuilt IVR feature that got used extensively [4, 5]. Our reason to look beyond community radio to instead use mobile phones was the slow licensing process to get a community radio broadcast license in India, and the restriction by law of the broadcast being constrained to only 10-15km. A service accessible via mobile phones could have a much wider outreach, and could be set up and scaled easily without licensing restrictions. We were also inspired by the success of a few other pioneering initiatives using IVR technology where people could call and listen to audio, or record their own voice messages that could be heard by other people [6, 7].

Our IVR initiative, *Mobile Vaani* (MV), runs in a similar manner: People give a missed-call to a unique phone number. The IVR system initiates a call back to the callers effectively making it a free service for them, and the people use phone keypad buttons to browse a list of audio messages or record their own message. Any messages recorded by them go through a manual moderation process before being published on the IVR for others to hear, essentially to filter out poor quality audio recordings or undesirable content.

Our goal with this initiative was to bring about social development through the use of technology, along the following five lines. First, we were convinced about the transparency and accountability function that such a platform could play in local governance, by making it simple for people to share about civic or governance issues that they were facing, and to demand transparency in decision making. Indeed, we had seen our community radio system in some locations used to run live broadcasts of Panchayat (village level elected local governance body) meetings that was much appreciated by the community [5]. Second, we had evidence from our own prior work in social media analysis [8], that discussions on participatory media platforms are specifically useful because the homophily effect [9] leads to people receiving information from their strong ties [10] which is highly contextual and understandable by them, and participation by diverse users from across weak ties leads to information that enhances completeness. We hypothesized that messages evolve as they traverse the social network, gaining both context and completeness through comments made on the messages by people occupying different positions in the social network graph [11]. We therefore believed that since MV could facilitate people to share their views and experiences on various topics, it could lead to a more complete and actionable understanding gained by the people for relevant topics that could include health or agriculture or education or potentially anything. Third, we had seen a strong community building impact emerge from community radio stations through traditional songs and music, and coverage of local events, which made the community closer knitted [5]. Fourth, we were convinced through our own field visits and prior work about the self-empowerment that people gain when they can be heard by many others, ie. when they gain a voice, and are able to challenge exploitative power structures [3]. Finally, we wanted MV to have an operations model and a business model that could be replicated readily for scaling either by us or through partners to bring wider and large-scale impact.

We therefore initiated MV with a purposive clarity of what we meant this platform to become, and the change we wanted it to bring about. Our approach was further driven by an assumption that these would be the exact impact pathways through which social development would arise through use of the MV technology. We were wrong, as we explain next by highlighting many different challenges we faced in realizing this vision, because we had not considered the socio-technological interface that shapes the use of technology. However, by building processes to manage this interface, we were finally able to reasonably achieve our vision. We consider this as a case for being able to successfully control the direction of technology use towards meeting specific objectives. We first discuss our own experiences and the processes we developed within the MV initiative, and then build an argument about possibly how similar ICT platforms could be controlled to lead to responsible outcomes.

III. THE SOCIO-TECHNOLOGICAL INTERFACE OF MOBILE VAANI

We define the socio-technological interface as the boundary with technology engineering and user interface design on one side, and factors that shape how people use the technology on the other side. We do not mean to suggest that people who engineer the technology or design the user interfaces do not consider the social context in which the technology is intended to be used. In fact, several of their choices may be shaped by the social context very much like how our choice for using IVR was shaped by the low literacy and high mobile phone usage in the communities of interest. The socio-technological interface however that we talk about is in the dimension of how the technology is used, and not how it is engineered. This understanding of technology use is typically considered to be the domain of social scientists. Partly by chance and partly by design, over time we expanded and diversified our team which had earlier been composed of only engineers, to build such a social sciences function internally, and converted MV almost into an action-research project where we eventually learned how to manage different aspects of the socio-technological interface to achieve our vision. Some of these aspects and the processes we developed to manage them are described next, and may generalize to other ICT initiatives operating in a similar context because the listed aspects are likely to be universal in the given context and similar challenges will be encountered by other initiatives as well. The listing of aspects we talk about are by no means exhaustive, we do not discuss privacy and security for example due to a lack of experience on this front, but nevertheless this list is intended to serve as a starting point to which more can be added over time. We also do not engage in a discussion about whether a particular technology is suited to meet a particular objective, our analysis is limited to managing how a given technology is used by society to meet its stated objectives.

*A. Technology and platform literacy, technology access*

We found that IVR itself was an alien concept for many people we interacted with. Our initial entry into the field in 2012 was through human rights activists working in rural areas of Ranchi and Khunti in Jharkhand, with people heavily dependent upon government welfare schemes such as NREGA (for rural employment) and PDS (for subsidized access to food grains). Men had access to phones and were functionally literate to use it to dial numbers and pick up calls, but most had not used automated IVR systems in the past. The usage of the system had to be demonstrated to them, and in the absence of any other media in these areas, offline training sessions were the most effective[1]. Further, they had to be explained the concept of community media and what they could do with access to such a platform, but that too was not straightforward to convey – it required practical examples and self-validation for people to understand the use of the technology. It is important to highlight that media penetration in these communities is extremely low, for newspapers due to literacy issues, and for television due to affordability and electricity infrastructure. Finally, access of phones by women, and access to women to tell them about MV, were both significant challenges. Due to patriarchal norms, women are typically less literate, do not own their own phones, rely on shared access, and consequently their capability to use phones is lower than that of men. Physical mobility of women is also lower and hence it is harder to reach women to participate on MV.

We had to develop low-cost and scalable processes to overcome these barriers of technology and platform literacy, and technology access. The wide usage of mobile phones was clearly not going to easily translate into MV adoption.

Related work has explored several interesting dynamics through which new technologies are learned. Poorly literate construction workers were able to learn a complex sequence of steps to share videos over Bluetooth, suggesting that self-motivation to use technology (in this case, for entertainment) could lead to self-learning [14]. In a study of Facebook use among urban youth [15], a mix of financial and social incentives led to users teaching their peers about the platform. In the context of women, [16] observes that despite strong patriarchal norms, women learn to navigate family and community spaces to learn how to use mobile phones.

We came across similar dynamics in MV as well. Users would tell their friends about it especially if their message got published on MV. Some would listen to MV in groups which led to wider listening and learning. Hearing stories of validation of MV's impact especially in the area of grievance redressal which we explain later, also led to MV achieving strong social credibility and popularity through word of mouth. These were however not systematic and reproducible processes that could be considered as part of a replicable MV model. We in fact made an attempt at formalizing these processes by giving a financial incentive to users to train other users, but it got misused and devolved into spamming as users falsely claimed having trained some people, but whose phone numbers had been acquired from mobile shops and spam databases. We therefore took a different approach, as explained next, to develop an innovative low-cost offline process through community volunteers which could manage this need to create technology and platform literacy.

*B. Community embeddedness and usage definitions*

As a community media platform, MV espoused a vision that its agenda should be driven by the community itself, including the moderation process, choice of content topics, etc. Related work has discussed the concept of *communitization of technology* [17], ie. when a community learns the essence of what a piece of technology can do and is able to leverage it for the community's needs. The study highlights the role that a few key technology savvy community members play in the process. These people are termed as Human Access Points (HAPs), and are essentially more technologically advanced users who understand both the needs of their community and the capabilities of the technology, to be able to realize relevant use-cases for the technology.

This notion of HAPs most closely explains our own methodology to *communitize* MV, and it happened organically. Through different partner organizations, as we were introduced into new communities, we kept coming across HAPs who quickly understood the technology and were among its early adopters. To gain faster popularity for MV in the community, we inducted people from among the HAPs as community volunteers, and built a financial incentive model to cover for out-of-pocket expenses incurred by the volunteers to popularize MV and guide its usage. They would travel to different villages and tell people about MV, demonstrate its usage, and encourage its adoption. We however realized that most of the MV volunteers may not be able to manage the time for content moderation themselves, plus due to their limited exposure to computers worsened with flaky Internet access in small towns and rural areas, we decided to retain the moderation function centrally within our team[2]. We encouraged the volunteers though to discover use-cases for the platform on their own, based on their understanding of the community [18]. As a result, over time when MV expanded to different locations running their own local MV chapters, the volunteers created their own topic priorities for respective local MV chapters. The volunteers in one location where MV was heavily popular with farmers, built linkages with the local agricultural institutes to answer questions put up by farmers. Another chapter built linkages with school and college coaching classes to advice their predominantly youth userbase with career counselling. Almost all the locations also took up a hyperlocal news reporting function that we discovered to be a universal need in all the MV areas of operation, understandably due to the scarce penetration of any other media.

We also supplemented the platform with content created by our team, on cultural and entertainment themes, discussions on social norms such as early marriage and domestic violence, rules and regulations behind different government schemes, etc. The choice of this content was guided strongly by feedback processes we developed to interview some regular users over the phone, conduct focused group discussions in the community, and also run IVR-based surveys to get user feedback [18]. In this way, even though we retained the moderation process centrally, we were able to achieve a good community embedding of MV through its volunteers and a process of continuous communication with them.

---

[1] In a different context and some years later we also ran experiments to show that offline learning was most effective to train users on IVR, as compared to delivering a training over a phone call, or broadcasting instructions over radio [13].

[2] It is only now as of 2018, more than six years after we started MV, that we feel Internet access is good enough and rich smartphone apps can be developed for content moderation instead of browser based content management, that we have started thinking again about building the originally envisioned distributed moderation model for MV.

Our journey with identifying and training volunteers, and retaining them, was however by no means smooth. We next discuss the question of internal accountability expected of the volunteers to build MV into a sustainable institution in itself.

### C. Sustainability and internal accountability

Initially MV started as a state-wide service in Jharkhand that was popularized in different areas by the volunteers. Subsequently, we started services also in Bihar and Madhya Pradesh. Our field team however consisted of only a few people who began to find it difficult to coordinate with dozens of volunteers from across different locations. Further, volunteer attrition became high because we discovered that many people would join MV as volunteers with the expectation of financial returns, but the small stipend we offered did not provide them with a strong monetary incentive. We realized that we had to improve our selection process to identify volunteers who were genuinely interested in bringing a positive change in their communities and for whom these social incentives would be stronger than the monetary incentives, and also build structures for the volunteers to be able to motivate and inspire each other. The concept of federated groups in the context of trade unions has been suggested as being more resilient than a single large group [19], and we adopted this method.

We split the state MVs into district level local chapters, each of which had its own unique phone number and could build its own identity [20]. Further, we grouped the volunteers from each district into a *volunteer club* for that district. The club elects a coordinator and meets on a monthly basis to discuss their activities and plan for the way forward. Through this hierarchical arrangement, our field team now only had to engage with the club coordinators. Further we found that this structure also built a strong solidarity and mutual accountability among the volunteers, and reduced attrition to practically zero. We also developed an elaborate financial incentive model as a combination of a group incentive that was divided equally among all the volunteers in the club (calculated pro-rata on the number of active users in each club) and individual incentives for each volunteer (based on the number of good quality contributions by the volunteer, and offline community mobilization activities organized by them). This model further made explicit the ethos that a volunteer club should act as a single unit to which all the volunteers were expected to make individual contributions to achieve the club's collective aim. The club model helped us realize the kernel for MV operations. It has already been tested with replication at thirty local chapters, and can be scaled readily for large-scale coverage and impact. However, as we explain next, this still left us with a few unanswered questions about the ideal volunteer profile to select and consequent processes for outreach by the volunteers.

### D. Appropriation and empowerment, inclusion and exclusion

The unique position of power occupied by the MV volunteers poses several risks. The volunteers can popularize to the users an altered vision of MV more relevant for them based on their individual socio-economic-political views and priorities which could be different from that of their clubs. They can similarly prioritize creating access for a select group of users by training them well while excluding others. It is also possible that the volunteers may have the best of intentions to facilitate equitable access but their ability may get impeded by socio-cultural norms they find hard to transcend themselves. We found all such dynamics taking place in practice.

As an example, during the initial days of MV, several volunteers were associated with human rights activist organizations and hence they were more interested in governance topics to the extent that they would discourage people from using MV for cultural expression through folk songs and poetry [20]. In another case, a class-based conflict arose in a club when a lower-class volunteer was elected as a club coordinator. Such incidents have now become rare due to the more rigorous selection and training methods for volunteers, but what was useful for the MV team to make the necessary course corrections was an openness to hear complaints that the users recorded on MV or shared with us in field meetings. This roughly designed internal grievance redressal process helped empower the users, and helped us to uncover such cases of undesirable appropriation of technology.

A challenge which still remains however is in overcoming the technology gender divide. Most of the MV volunteers are men and find it hard to reach women to tell them about MV. Having a female volunteer in a club of all male volunteers is also difficult in the dominant patriarchal cultural setup of rural India. An all-women MV club was also started and despite all the volunteers being extremely dynamic, the active userbase of the club remained low due to the limited mobility of the women volunteers to reach other women. It is worth mentioning that in a recent project we worked with a large women Self Help Group (SHG) network and were able to reach many women through SHG meetings [21]. The regularity of the SHG meetings which take place for financial bookkeeping, and the exclusive women constituency to which we gained access, provided us with an opportunity for both targeted outreach to women as well as repeated interactions with them to encourage technology adoption. Gaps still remain though. For instance, meetings with SHGs of very poor women were held irregularly since the women were busy with work and hence those who could potentially benefit the most from access to the platform were excluded even via this pathway.

To summarize, we succeeded although not completely to manage the socio-technological interface in MV about biases in inclusion and exclusion that arose due to characteristics of the social context, or even due to appropriation of the technology by those who were more powerful and adept users of the technology. Not tackling this challenge stands the risk of increasing the inequalities that exist in the social context, because the more powerful or adept users of technology are able to leverage it for their agenda and move further ahead, leaving the others to play a perpetual catch-up game [22].

*E. Social and institutional credibility*

It is reasonable to expect that sustained participation and utilization of a technology platform will only happen once the expectation of the users is met. MV is popularized as a community media platform with several characteristics and use-cases pitched to the users. Some of the significant examples include: "*It is a platform for you to talk about whatever you feel is relevant for you and your community*", "*It is to highlight news that is relevant for you in your daily life, and not what is sensational*", "*You can get breaking local news way before any newspaper or TV channel*", "*You can discuss local and national policy, and we will convey your feedback to the right stakeholders*", "*MV is a platform where you will get useful information on agriculture, career counselling, health, government schemes, among other topics*", "*MV volunteers will help resolve problems that your community is facing, especially on welfare entitlements and public services*", etc.

We would like to explain the context of the users so that the importance of meeting these expectations can be better understood. Regional newspaper and television media has not deeply reached these communities, and also has a chequered reputation of having ignored problems of lower caste people, or having suppressed news against the local elite possibly even in return for payment. The poor and marginalized groups have historically lacked a strong voice in the community, and even their political representation has had its own ups and downs. They are also intimidated by complicated government office processes for grievance redressal, and may not be able to take time off their daily livelihood routine to pursue even legitimate cases of violation of their rights and entitlements. In such a context, when MV is positioned as making several strong promises to the extent that it will enable the people to overcome the social inequalities in which they live their daily lives, then if the expectations of the users are not met they will dismiss it cynically as yet another fake promise.

MV was able to gain significant social credibility by showing evidence that helped people validate its stated promises and intentions. The editorial processes of moderation have been kept liberal, and only filter out poor quality audio messages, or those that appear to be politically motivated or are spoken in a rude tone. Towards the initial stages of MV, the moderators would even make phone calls to users who seemed to be wanting to say something important but were not able to articulate it well, and guided them to be able to make better audio recordings [3]. Similarly, grievances recorded by the people, or questions asked by them, are rigorously followed-up by the volunteers with reminders and support provided by the moderators, to be able to convey to users about action having been initiated upon their requests. All the MV volunteers are also trained on news reporting to cover news events that they come across. These functions of grievance follow-up and news reporting, in addition to informing users on technology and platform literacy about MV, form a significant part of the offline work done within the MV ecosystem by the volunteers to ensure strong social outcomes.

A recent survey of several hundred MV users showed that 67% of the users agreed that it is different from other mass media in giving an opportunity for anybody to voice themselves, 69% acknowledged the value of dialogue created on the platform to understand different viewpoints, 88% reported an increase in connecting back to their cultural roots, 85% reported an increase in political awareness, 50% acknowledged having learned new ways to articulate their views, 64% reported having gained agency in addressing problems with local governance directly themselves, and 84% acknowledged strong offline support received from MV volunteers in helping solve their problems.

A categorization of different impact pathways of MV is provided in [23], constituted on the basis of close to a hundred indepth interviews of users, on different types of learning and agency effects experienced by them. A further detailed discussion of social accountability loops created by MV is provided in [24], based on 300+ impact stories shared by the users on MV. The argument is generalized in [25] through a collection of case-studies of IVR-based projects which were used to create new information flows that brought transparency and empowered users to demand better quality in the delivery of public services. [24, 25] also emphasize that the target beneficiaries of welfare schemes are much in need of offline support via social workers to demand their rights and entitlements because self-service mechanisms like centralized helplines and web portals can end up being disempowering as they are hard for the people to use. Furthermore many categories of grievances arise due to issues at the local level and cannot be resolved through centralized mechanisms. MV's network of volunteers not only provided such offline support to overcome the disempowerment of the people in engaging with the government authorities, but are also able to use MV to channelize the attention of government officials towards specific grievances and bring about high rates of redressal.

To summarize, the platform processes of moderation and editorial policies, and offline processes by the volunteers of user outreach, support, and follow-up on user requests, were able to meet user expectations and were critical for MV to gain social credibility. This in turn created other effects to facilitate strong word-of-mouth popularization of MV by users who benefited from it and readily endorsed it.

Institutional credibility too is important for MV to be able to get constructive reactions from the state for grievance redressal or policy implementation feedback given by the users. As the MV volunteers built stronger networks with local government officials, and demonstrated sustained usage over many years, we found that the state too responded positively and viewed MV as an innovative means of citizen engagement that they could utilize themselves. Local government officials now routinely use MV to make announcements of new schemes and subsidies, give interviews about their views on policy implementation, offer a commitment for resolution of grave issues, and respond actively to requests by MV users and volunteers. Institutional credibility thus reinforced social credibility, and helped embed MV not just with the community but with other local institutions as well.

---

[3] These manual guidance calls by the moderators were later discontinued when the userbase grew and enough resources were not available to sustain this function. Currently, an automated call is made whenever an audio message is rejected from publication, and pre-recorded tips are provided on how to make good quality recordings. Going forward as the MV distributed moderation function shapes up, we plan to reinitiate guidance calls to be made by the volunteers.

*F. Influence of the business model on agenda*

So far we have explained how we managed many aspects going beyond the technology of MV into its interface with the community and local institutions, and we have built an operations model that can be replicated and scaled. The final aspect we discuss next is the MV business model, so that MV can be scaled and made financially sustainable.

We have three revenue streams as part of the MV business model. First, philanthropic donations and government advertising to fund awareness and behaviour change campaigns on topics such as health, nutrition, education, and livelihood. Second, community funding where the users may themselves contribute small amounts for their community media platform, plus crowd-funding to raise micro-grants for specific activities and sponsorships from economically well-off people. Third, commercial advertising by companies interested in reaching rural markets. While there is strong validation of the first revenue stream, the second is completely untested as of now, and the third is yet to be tested at scale.

We believe that all three revenue streams will kick-in sooner or later, be it on IVR or on a future mobile application running on data services, since there is a lack of appropriate media outreach channels for Indian rural markets. What we do not know however is what challenges will come up in sustaining the unbiased and community driven coverage provided on MV currently. As the revenue streams from government or corporate advertising get larger and MV's dependence on them increases to ensure its financial sustainability, MV is likely to become susceptible to have its agenda get influenced by government and corporate interests. We do not have any experience so far in building processes to manage this likely forthcoming challenge, since MV has until now been sustained either through philanthropic grants or internal funding by Gram Vaani through positive margins made on other services.

IV. MANAGING THE SOCIO-TECHNOLOGICAL INTERFACE

We showed in the previous section how the MV team diversified from its core expertise in technology engineering to take on a more action research oriented approach towards using technology for social development. Our key objective with giving a detailed description of our experiences was to demonstrate the complexity of problems that had to be understood and solved to achieve a reasonable degree of success. We showed that indeed the use of technology for development is complex, but it is possible to build processes and ensure that the technologies can be controlled towards leading to responsible outcomes. The very same problems are likely to arise for other ICT platforms as well, which will hit upon similar challenges and will have to manage the same socio-technological interfaces. Our goal in this section is therefore to attempt to generalize the processes developed in the MV context so that they can be incorporated in other technology initiatives as well, where ICT platforms are used by the end-users directly to share information, especially towards achieving certain social development outcomes. Not being able to manage the interface between technology and society stands the chance of technologies that were meant to be empowering for people, to actually disempower them, or some of them.

Distilling from the MV experience, the following specific processes were employed to manage its socio-technological interface. The processes are listed as headings in the enumeration below and the aspects of the socio-technological interface they specifically help address are underlined in the description alongside.

*1) Federated setups:* These make management of the ICT platforms easier and allow for both contextualization and diversity in the use of the technologies. Smaller communities can build specific <u>usage definitions</u> according to their needs, which can also lead to greater <u>community embeddedness</u> of the technology, rather than expect all users to have a uniform worldview about it. This embeddedness may be brought about through a subset of users who can mediate as community representatives, either online or offline.

*2) Internal feedback processes*: Platform providers can get regular feedback about the platform usage from users on aspects such as the following.

   *a) Grievance redressal:* This can alert about cases of <u>misappropriation of the technology</u> by malicious users, or <u>malfunctioning of the technology</u> itself, both of which are likely to happen in any ICT platform. Being able to address these issues promptly can contribute towards <u>social credibility</u> of the platform in the eyes of its users, and also <u>empower</u> and <u>incentivize</u> them to have the ICT platforms not get misused.

   *b) Tracking of inclusion and exclusion:* Biases in access or usage arising from gender or other categories of inequity can be spotted, the reasons behind the biases can be identified, and appropriate action can be taken. This can help avoid an increase in inequalities through <u>inclusion and exclusion</u> biases arising from inequities in the social context itself that constrains access to the technology for different usergroups.

   *c) User interests and context:* ICT initiatives can learn more about the interests of their users to be able to build relevant <u>usage definitions</u> leading to more appropriate products and services. Most platforms do this extensively through implicit learning of user preferences which is also an issue of debate on privacy as ICTs facilitate access to detailed data about user actions. Platform providers however rarely engage in getting closer to the users to understand them, especially if the users inhabit a different context like being situated in a different country or being from poorer backgrounds, and the platform providers do not get explicit feedback from the users that can help shape the service directly towards specific social development goals.

*3) Design of incentives:* Managing federated setups and building a closer feedback loop with the end-users may require more efforts and hence higher resource costs for the platform providers. Appropriate incentives can therefore be built to involve the users themselves or their community representatives in the management of the platform, like in managing federated setups or providing feedback, ultimately costing much less. Suitable social, solidarity, and monetary initiatives can lead to strong sustainability and internal accountability outputs, and create long standing community embeddedness. Further, being able to align the incentives with positive social change as an overarching objective may lead to greater participation by the users towards platform management and build its social credibility.

*4) Processes to address gaps in technology and platform literacy:* Not all users can be expected to have a good understanding of how to use the technology, and what it could be used for. Tracking inclusion and exclusion metrics, and usage profiles, through the feedback processes mentioned above, can help spot specific biases that could be arising from this disparity in technology and service literacy. Appropriate steps can then be taken to bridge the gaps. Depending on the context, this gap could be bridged through online mechanisms or it may require offline mechanisms as well, and it may be undertaken by the institution administering the ICT platform or it may require incentive mechanisms for the users or community representatives to participate. Sometimes bringing about this technology and service literacy may not be straightforward or even possible, and in such cases processes to facilitate assisted usage may be needed. All such steps can significantly avoid disempowerment effects that some users may notice from their inability to learn to use new technologies easily.

*5) Offline processes to meet the stated social development objectives:* This is especially important for ICT platforms aimed at achieving some promised social development outcomes. This is likely to need offline processes, well aligned incentives for the participating users and community representatives, and institutional linkages with external stakeholders, over and above the capabilities of the technology itself. Even if simpler outcomes are promised to the users, it is likely that the technology alone may not be sufficient and the ICT initiative will still need to build operations beyond the engineering of the technology alone. If the platform operations are able to establish a clear link to the expected outcomes, it will lead to the platform achieving high social and institutional credibility. Further, the linkages between the platform and external stakeholders should be such that they empower the users: over-centralization and automation has been seen to result in disempowerment effects when people are not able to utilize the links due to various social contextual factors.

We argue that ICT initiatives should embody these processes to be able to manage the socio-technological interface for their innovations. To summarize, this interface in the given social context to build ICT platforms for poor and marginalized users in India, is defined through its aspects to address societal gaps in technology literacy, platform literacy, and technology access; build community embeddedness and clear usage definitions for the people intended to use the ICTs; achieve sustainability and build processes for internal accountability; guard against technology appropriation by a few users, which especially in conjunction with societal biases can lead to biases in the inclusion and exclusion of certain usergroups, and in turn lead to greater inequalities; achieve strong social and institutional credibility by meeting expectations of the users; and create a business model that can bring financial sustainability but not compromise processes required to manage the rest of the aspects. Other ICT platforms in a similar social context are likely to stare at the same socio-technological interface and may need similar processes. We next describe a few other ICT platforms and the fallouts if strong processes are not built to manage the interface.

## V. Towards controlling technology

Now that we identified several aspects of the socio-technological interface that lead to undesirable outcomes from the use of technology, and processes that can be developed to manage this interface, we next move to the question of how can it be ensured that the technology providers put efforts towards managing this interface? Further, can realizing social development outcomes from technology be made a paramount objective of technology development itself? We start with laying out problems noticed with some popular technology platforms in managing their socio-technological interfaces, we then describe similar problems seen with some technology platforms led by the state, discuss the broader political economy which influences the priorities of market and state driven technology platform providers, and finally give examples of corporate governance structures that may help ensure that technology providers pay due attention to managing their socio-technological interfaces.

### A. Laissez-faire market driven technology expansion

Our approach with MV lies in sharp contrast with how ICT initiatives otherwise typically interface with society, placing emphasis largely on the technology engineering for achieving popularity and scalability, without adequate care about managing its socio-technological interface. The first example that comes to mind is Facebook which has lost considerable social and institutional credibility over recent years for the limited attention it paid to check the presence of echo chambers and filter bubbles created through its operating models and the metrics it chose to optimize, slow efforts to detect fake news when malicious users appropriated it for various political and monetary gains, ability to control data leaks, etc. Mis-appropriation happened even with Digg, even though it was conceived as a democratic space to counter the rich-get-richer phenomena on search engines and blogs, and it gave strong agency to the users to select content, but yet a few users were able to subvert these processes for political gains. Platforms like Reddit and Slashdot on the other hand, have so far shown resilience to such cases of appropriation. They have a strong moderation system mediated by people, and Reddit in fact is set up as a federated system which allows different communities to build their own standards. Other extremely successful collaborative knowledge building platforms like Wikipedia (and even Quora to an extent) similarly have strong moderation policies and a somewhat implicit federated structure defined mostly on the basis of topic interests of users. Facebook however it seems has only lately scaled centralized human driven moderation processes, and AI to detect problems, but its architecture is inherently not geared towards empowering users to take over the administration of diverse federated community groups, or at least is not used effectively. Rather the push seems to be in the direction of building universal community standards across all users of Facebook.

This effort to bring standardization in an inherently diverse society seems at least to us an exercise doomed to failure by definition. We argue that first, when such ICT platforms are deployed for use by people, people will undoubtedly use them for different agenda and consequently management will become an issue, therefore it is important to design the architecture of the ICT platform upfront with the acknowledgement that the socio-technological interface of the platform will need to be managed towards responsible outcomes, and that this is an inescapable eventuality. Reddit and Slashdot seem to be strong success stories here in having thought through the architecture design beforehand itself. Second, it is entirely understandable that even after sufficient forethought, this interface may still throw up unforeseen aspects. New processes will then need to be developed to manage them, and therefore it is again the prerogative of the ICT platform providers to acknowledge their responsibility and understand this interface to be able to manage it. Even though platforms like Facebook may not any explicitly stated social development goals, but it is clearly irresponsible to leave it to external investigators and researchers to point out societal problems that could be arising from their platforms, while they simply focus on scaling their platforms and scientifically optimize the usage targeting self-serving metrics only.

This brings us to WhatsApp. The WhatsApp architecture of allowing users to create their own groups and managing them, is closer to a federated and user driven setup, although we do feel that the tools available to the administrators are primitive and a lot more can be done. However, the encrypted nature of communication makes it practically impossible for anybody to manage it, and WhatsApp has nearly absolved itself of any responsibility towards facilitating more appropriate administration of its forums. Only now in India has it started holding workshops on fake news, but which remain at a minuscule scale; other methods even through online tools to create better technology literacy could be far more effective. Laissez-faire market driven technology expansion does not seem to have paid much attention either to understanding the suitability of different architectures, or to aim for responsible metrics and user incentive structures instead of those that link towards profit and shareholder-value maximization, or to ensure better literacy of its users towards the technology and services. This ideology of market driven rapid growth has undoubtedly enabled communication and collaboration at unprecedented scale, and which has been used for strong social development outcomes as well, but it is the missing element of acknowledging the need to manage the socio-technological interface that is our primary criticism. Not being able to manage this interface leads to many undesirable consequences, as has been noticed with so many large ICT platforms including Facebook, Twitter, WhatsApp, etc.

These problems are not constrained just to the kind of information sharing ICT platforms we primarily discuss here. AI technologies are also criticised for their bias, often arising due to biases in the training datasets and the absence of feedback loops about when the predictions of AI systems make mistakes [41]. Several efforts are underway to bring fairness, accountability, and transparency in AI systems [42], but we believe that even with these technological modifications, there will be a need to go beyond just the technology engineering to manage different aspects of the socio-technological interface, especially those dealing with technology literacy so that users are made aware of possible limitations of the technologies and remain alert enough to spotting issues which they can raise through suitable internal grievance redressal processes.

We turn next to the question of how it can be ensured that market-driven ICT platform providers manage the socio-technological interface, after first a discussion of gaps in state driven technology expansion, and a possible explanation of the political economy that seems to be driving the market and the state in their methods of technology expansion.

*B. State driven technology expansion*

The state needs to hold itself up to even higher standards of responsibility in the technologies it chooses to use. This is justified since the technologies are likely to be funded through tax-payer money, and may get deployed at very large scales, even possibly through coercion by mandating citizens to use them. We take the example of Aadhaar, which is not exactly an ICT platform meant for end-users to communicate with each other, but a significant use-case on the platform is for end-users especially the poor to execute welfare related transactions with small and large businesses, which exposes Aadhaar to similar challenging aspects of the socio-technological interface.

Aadhaar is arguably a fitting demonstration of the high-modernism ideology of the state to bring simplification and legibility of its population towards being able to administer them more easily [26]. It has allegedly been instated through unconstitutional means, has been pushed through coercion by the state, and driven for straight mass adoption without adequate testing and piloting of the technology. This has led to many issues especially grave problems faced by the poor, which clearly could have been avoided had more attention been paid to the socio-technological interface. Better user feedback processes could have been built. Campaigns to bring about better technology and service literacy could have been conducted. The capability of the users and their context could have been understood better, to support them with mechanisms like offline linkages for assistance instead of building over-centralization and self-service mechanisms which end up being disempowering for them. Alternate methods to conduct transactions instead of Aadhaar could have been allowed, especially given that unforeseen problems could arise in a mass rollout of such new technology. In fact, leave aside the acknowledgement of the presence of unknowns in the socio-technological interface, the technology engineering itself has had issues that made Aadhaar seem like an irresponsible move without adequate testing for robustness of the technology working in the wild.

Further, it is debatable whether Aadhaar was the right choice of a system to meet the stated objectives of reducing corruption and leakages in welfare benefits[4]. Other systems that empower users by bringing more transparency and accountability in welfare services provisioning, have been seen to work and could have been good alternatives to Aadhaar [44]. The alternative we are alluding to here is a village-level SMS alert system piloted in Uttar Pradesh that would push SMSes to the recipients of subsidized food in the village, whenever a new delivery of grain, kerosene, and other food products was made to the fair price shop in their village. Armed with this information, people from the village were then able to evade tactics of the shopkeeper to withhold food under the pretext of not having received the required quota of products from the government, which the shopkeeper would siphon off to private channels of distribution. Such a system was empowering for the community and community based institutions, and built up their confidence in demanding their rights and entitlements for public services in general. The Aadhaar method however resorted to the use of biometrics which failed in the field, problems arose due to incorrect details filled in carelessly by enrolment agencies that could not be corrected through any easy means, disempowered the users not equipped to handle complex technologies and protocols like OTP verifications, and disempowered the community institutions who were rendered helpless to help the communities.

Policies like demonetization also bring out the arbitrariness of the state in policy formation and its capability for coercion towards compliance, hence on the one hand it is not altogether surprising that ICT policies like Aadhaar were handled similarly by the state, but on the other hand it speaks about the general attitude of the state which in this case can be argued to be unethical on all of political, moral, and scientific grounds.

*C. Contemporary political economy around ICTs*

What can explain these dominant ideologies for market and state driven technologies to achieve massive scales yet not pay adequate attention to managing the socio-technological interfaces? It is not hard to understand the market rationale for scale, bring driven by surplus financial capital looking for investments, and the ability of ICTs to transcend geographical barriers that sometimes limit other industry sectors from catering to a global market. Such investments only eye profit and shareholder-value maximization, always aimed at optimizing only performance metrics that lead to financial gain. Achieving social development gains is not a priority for such institutions, it is only an incidental outcome. Their only priority is to not violate any laws or regulatory requirements in their pursuit of profit, and also make efforts to influence the regulation through strong lobby networks with the state. This is the part we want to touch upon, and also from the other side, of the state's willingness to associate with them.

---

[4] We want to clarify that the appropriateness of system design is not the primary focus of this essay, rather it has been framed in terms of managing the socio-technological interface which arises after a given system has been designed. We regardless do digress somewhat in the context of Aadhaar to be able to present a more complete discussion.

The state may have at least three incentives to associate with large ICT providers. First, the state's ideology of high-modernism can benefit from the surveillance potential of ICT platforms, or use these ICT platforms for its own propaganda. Second, especially in democratic setups the politicians are able to make promises to citizens of easy change through rapidly scalable technology, and are assisted by media to a significant extent, as we explain later, in building up the hype of benefits that can emerge from the technologies. Third, allowing the platforms to operate with light regulation, and creating new spaces for them such as the large industrial ecosystem that has emerged around Aadhaar, may provide the state with new rent-seeking opportunities. This complex of interlocking ideologies between capital and the state that allows both to pursue their respective goals, has led to rising inequalities [27, 28], we have also found that the interlocks are getting stronger over time [29], they influence policy [30, 49], and the mass media itself has been appropriated, possibly through choice rather than coercion, which limits its ability to influence public opinion on appropriate policies [31].

Even if public opinion does become strong to the extent that in democratic setups it can overcome the power of state-capital interlocks and force the state to both act responsibly on its technologies and to also exercise regulation on market players for technologies being pushed by them, it raises a concern about whether regulation can indeed keep pace with that of technological development? It seems we do not have strong evidence so far about this possibility, or even to define frameworks in which to think about regulation. This probably leaves one other factor to shape technological development along more responsible lines, competition, but which too seems to be compromised due to the near global monopoly enjoyed by most ICT platforms in wide use today. In fact, the tendency of capital to centralize itself actually drives industries towards less competition [47], and it is even arguable whether new technology disruptions can be assumed to arise and reset the balance in the future [48].

This brings up critical questions that even in such a market-state political economy of today, can market driven ICT platform providers be made to pay closer attention to managing the socio-technological interfaces touched by their platforms to avoid negative outcomes? Can they be made to aim for social development outcomes rather than financial outcomes? Can the state also be made to manage well the socio-technological interfaces for platforms adopted by it? We next outline some possible directions which may help move towards these goals.

*D. Mastering the social control of technology*

There are differing perspectives about the interaction between technology and society. The theory of technology determinism suggests that technology and its development follows its own logic and ultimately shapes society, leading to eventual progress [32]. The theory of social shaping of technology suggests instead that society shapes what technology gets created and how it gets used [33]. The most relevant to our discussion is however Collingridge's analysis of the question: "*can we control our technology – can we get it to do what we want and can we avoid its unwelcome consequences*" [34]. He raises an important dilemma that when a technology is young and can be controlled then too little is known about its harmful consequences, but when the technology gets widespread then it becomes hard to control because of its entrenchment which increases the costs for change, increases debate, and even gives the status quo an unfair advantage to argue against change. Decisions about controlling the technology therefore may get shaped by many social and political forces, and may often need to be made without complete knowledge. Our effort in this essay has been to highlight that rather than make decisions in the dark to shape the technology, many aspects of the socio-technological interface that lead to harmful consequences are already known and if the technologies are built around architectures that embody the kind of processes listed above, it may well be possible to make informed decisions and manage the socio-technological interface towards responsible outcomes.

However, as discussed in the previous section, it seems that the regulatory potential of both markets and states is limited to ensure that adequate attention is paid to following such due-diligence processes to control technology. This leaves the people – the consumers and producers of these ICT platforms – to redefine the priorities of the platforms. This we feel needs be done on two fronts, to ensure that the socio-technological interfaces of the technologies used or built by them are managed responsibly, and the technologies lead to social development outcomes for the poor and marginalized. We believe that there is enough talent in the world to address the first goal of responsible management of the socio-technological interface, it is just that it needs to be done without exception. The technology engineers and researchers whose labour is producing these technologies, can redefine internal organizational priorities to meet this goal. Likewise, consumers of the technologies can also shape the direction of use given suitable technology architectures that empower the users, and given sufficient consumer awareness to be able to exercise their influence. The second goal of working towards the needs of the poor and marginalized is, we believe, the only worthwhile goal so that the poor can equally benefit from technologies that have serviced the privileged, and define what new technologies should be developed, otherwise inequalities will continue to expand because the underserved will be left trying to catch-up to levels that continually become harder to reach since the privileged will be able to build and use technologies for themselves to advance further ahead [35]. Historically, the state and market have significantly shaped what technology gets built and marketed, and who the intended user is, why then for once cannot we build technology with the poor and marginalized as its primary users?

Such a redefinition of priorities by the workers was attempted as part of the Lucas Plan in 1976 in UK [36]. Lucas Aerospace was a company involved in the manufacture of aerospace components, and was considering restructuring and layoffs when the workers put together a plan of how the same aerospace technologies of the company could also be put to use for socially beneficial products, and save jobs too. The workers followed what is today known as a human centred design approach to conceive these products – they visited hospitals, old-age homes, etc to come up with designs, conducted market research exercises to understand the commercial potential, and finally recommended a plan to the management. This plan included products like vehicles for mobility of physically disabled people, portable life-support systems that could be used for medical emergencies, etc, and some were even built and used. Although the plan was eventually rejected by the company, the Lucas Plan is referred even today as a model of how the workers took control of what they wanted to do with the technology they built. Mike Cooley who led the Lucas Plan movement, in his book *Architect or Bee*, refers to a comparison that Marx drew between an architect and a bee, where "*the architect is able to imagine a plan and then erect it through his labour by subordinating his will to his plan, which helps the architect achieve consonance with his purpose*" [37].

How can producers of a technology be empowered to be able to determine what the technology gets used for, and ensure that the socio-technological interfaces are managed well to lead to the intended outcomes? There is reasonable precedence for some methods. German corporate boards have a notion of co-determination where worker representatives, typically from recognized trade unions, are able to participate in the management of their companies [38]. They are able to shape working conditions and economic decisions of the company, and could become conduits to channel worker preferences about technology use as well. The same idea of creating diversity in the board representation to improve corporate governance could be extended to also include social scientists in the board, who can ensure that the companies pay attention to the socio-technological interface. Another proposal suggested in [45] is to revisit a Code of Ethics for engineers that was originally formulated by IEEE in the 1970s, emphasizing upon disclosure and whistleblowing, and rejecting the assumption that the market could guarantee social outcomes arising from technology. The social development sector also has a lot to offer to deal with the complexities of technology use, having understood since long that technologies or any interventions do not operate in a vacuum, and are influenced by numerous social-cultural-economic aspects that need to be managed as part of an overall programme design targeted at specific outcomes [39]. They have developed systematic methods to conceive theories of change, specify the indicators to measure, the measurement methods, and ongoing feedback processes for concurrent monitoring to be able to take informed decisions for any corrections in the programme implementation [40]. Finally, the methods to identify relevant problems to solve and how, have also seen systemization [43], and hundreds of innovation incubators around the world offer guidance to entrepreneurs and innovators to spot tough social challenges and figure out how to solve them.

Assuming suitable governance structures can be formed to ensure that technology providers responsibly manage the socio-technological interfaces, there however remains an inherent risk for market-based enterprises to attempt to go beyond the technology engineering. It is expensive to do so, probably more error prone and risky than the engineering itself, and more complex simply because of the larger number of variables that may influence the outcomes. This can impact the financial sustainability and business model of the enterprises. Even if some new financing methods like social bonds are used to finance the operations, it leaves unanswered the question of how to price the investments made in the enterprises. The platforms of competing enterprises may also differ in complexity and hence their cost, but it is not necessary that complex platforms which might be more expensive to operate will also lead to stronger social development outcomes. Measuring the outcomes may not be straightforward either. These are important questions for different communities of research and practice to engage with, and define the way forward, possibly through suitable tax structures or even in terms of alternate institutional structures such as platform cooperatives [46] that can compete with yet operate differently from typical market based enterprises of today.

As technology gets more complex and driven by skilled manpower who are given space for their creativity arguably unlike the Marxist proletariats[5], we feel the answer may lie in realigning the relationship between the technology providers, the producers working within the provider organizations, and the users of the technology. Of course, for this to work, the technology producers will need to step out from their core engineering roles to interact with users of the technologies they have produced, to understand their context and close the distance between themselves and the users.

## VI. CONCLUSIONS

As new technological development takes place at a rapid pace, it becomes important to learn how to control the technologies so that their intended outcomes can be achieved and the technologies can serve towards social development of the poor and marginalized, empower them, and reduce inequalities. We have described several processes operating at the interface of technology and society that can help manage the outcomes arising from the technology, especially for ICT platforms which enable end-users to share information with one another. We also highlight the need for suitable corporate governance mechanisms to ensure that such processes are implemented by the ICT platform providers. This essay is also to emphasize to technology researchers and engineers that they should look beyond their role in just technology design, to understand its interface with society, persuade their organizations to ensure that responsible outcome arises from the technologies developed through their labour, and redefine what technologies should be developed in the first place by situating the poor and marginalized as the primary users of these technologies. Only then will technology really enable societies to overcome the inequalities that exist all around us, and not increase these inequalities further. All of us can together shape such a future.

---

[5] We acknowledge that whether or not the white-collar workers of today have more agency than the blue-collar workers, is a debatable topic. Writings of Harry Braverman, Charles Wright Mills, and Emil Lederer appear to suggest otherwise that even the political ideologies of the two groups have diverged. However, recent employee protests at companies like Facebook and Google over the unethical use of technology, and even gender equality at the workplace, were successful in drawing a responsible response from the management, and may indicate greater agency in the hands of the workers. We remain optimistic and strongly believe that educational institutions need to nurture this awareness right from the start in the engineers they produce. We are attempting to put together a course structure on ethics in applied CS towards this goal [50].